\documentclass[fleqn,usenatbib]{mnras}

\usepackage{newtxtext,newtxmath}

\usepackage[T1]{fontenc}

\DeclareRobustCommand{\VAN}[3]{#2}
\let\VANthebibliography\thebibliography
\def\thebibliography{\DeclareRobustCommand{\VAN}[3]{##3}\VANthebibliography}

\usepackage{graphicx}	
\usepackage{amsmath}  
\usepackage{rotating}
\usepackage{lineno}
\usepackage{multirow}

\title[Evolution of Time-Lags of Swift J1727.8-1613]{Evolution of Time-Lags of Swift J1727.8-1613 during the Rising Phase of Its Discovery Outburst}

\author[Nath et al.]{
Sujoy Kumar Nath,$^{1,2}$\thanks{E-mail: sujoynath0007@gmail.com}
Dipak Debnath,$^{2,3}$
Hsiang-Kuang Chang$^{2,4}$
\\
$^{1}$Indian Center for Space Physics,  466 Barakhola, Netai Nagar, Kolkata 700099, India\\
$^{2}$Institute of Astronomy, National Tsing Hua University, Hsinchu 300044, Taiwan\\
$^{3}$Institute of Astronomy Space and Earth Science, P 177, CIT Road, Scheme 7m, Kolkata 700054, India\\
$^{4}$Department of Physics, National Tsing Hua University, Hsinchu 300044, Taiwan
}

\date{Accepted XXX. Received YYY; in original form ZZZ}

\pubyear{\the\year{}}

\begin{document}
\label{firstpage}
\pagerange{\pageref{firstpage}--\pageref{lastpage}}
\maketitle

\begin{abstract}

We investigate the accretion dynamics of the black hole X-ray binary Swift J1727.8-1613 during its $2023-2024$ discovery outburst that lasted for $\sim10$\,months.
{\it Insight}-HXMT monitored the rising phase of the outburst of Swift J1727.8-1613 roughly continuously from 2023 Aug 25 to 2023 Oct 05.
Strong signatures of type-C Quasi-Periodic Oscillations (QPOs) are observed during this phase of the outburst.
In our recent paper, nature of the QPOs are studied with the propagating oscillatory shock (POS) model.
In this paper, we report on the observation of both positive (or hard) and negative (or soft) time-lags in the $4$--$10$\,keV (LE), $10$--$30$\,keV (ME), and $30$--$150$\,keV (HE) bands, computed with respect to the $2$--$4$\,keV reference band. We detect a clear transition from hard to soft lags as the outburst evolves. We show the evolution of QPOs and associated time-lags between different X-ray energy bands, correlated 
with changes in the QPO frequency, spectral state, and the size of the Comptonizing region. Our analysis reveals strong anti-correlations 
between the time-lags and both QPO frequency and photon index, and a strong positive correlation with the shock location.
These evolving lag characteristics and their correlations provide crucial insights into the changing accretion geometry and the interplay 
of radiative processes, further supporting dynamic models like the POS in explaining the coupled spectro-temporal evolution in black 
hole X-ray binaries.

\end{abstract}

\begin{keywords}
X-Rays:binaries -- stars: individual: (Swift J1727.8-1613) -- stars:black holes -- accretion:accretion discs -- radiation:dynamics
\end{keywords}



\section{Introduction}

Black hole X-ray binaries (BHXRBs) represent one of the most fascinating classes of astrophysical systems, providing a unique laboratory for 
studying the physics of accretion, relativistic jets, and strong gravitational fields. These systems consist of a stellar-mass black hole 
accreting matter from a companion star via an accretion disk \citep{SS73}. The infalling matter loses gravitational potential 
energy, which makes the temperature of the matter so high that it produces X-ray emission. This X-ray emission from BHXRBs show variabilities 
in longer (few weeks to months) as well as shorter timescales \citep[milliseconds to seconds; see][]{MC06}. BHXRBs are generally categorized as persistent 
or transient sources based on their long term variability \citep{Chen97, Tetarenko16, Corral-Santana16}. 
Transient BHXRB sources stay dormant with very low X-ray luminosity \citep[$L_X \sim 10^{30-33}$ erg\,s$^{-1}$;][]{Tetarenko16} most of the time, which is known as 
the quiescent state. Occasionally luminosity of these sources increases more than 4-7  orders \citep[$L_X \sim 10^{37-38}$ erg\,s$^{-1}$;][]{Tanaka96} 
for a period of a few weeks to a few months, which is known as the outburst state. 
At the start of an outburst, the source luminosity starts to increase rapidly. The energy spectra is initially dominated by non-thermal emissions, 
and the source is said to be in the hard state (HS) in this period. Gradually thermal disk emission starts to increase, the non-thermal emissions 
decrease and the luminosity continues to increase through an intermediate state. This state can also further be divided into the hard intermediate 
state (HIMS) and soft intermediate state (SIMS). Eventually, the luminosity reaches a peak value, and the spectra becomes dominated with thermal 
disk emission in the soft state (SS) \citep[see e.g.][]{Done07, Belloni05, Belloni11, Nandi12}. After that, the luminosity decreases to the quiescence level over a longer period of time, and the source 
follows an opposite sequence of state transition as the spectra become gradually harder. However, sometimes the softer and/or the intermediate 
states are absent in the so called `hard-only' or `failed' type of outbursts \citep{Tetarenko16, D17, Alabarta21}, and sometimes the X-ray 
luminosity shows more than one peak with a different sequence of state transitions \citep{Zhang19,Nath23,Nath24}.

Besides this long-term variability, BHXRBs also exhibit rapid variability in X-ray luminosity, which is typically studied by generating and analyzing the power density spectrum \citep[PDS][]{van89} from short-timescale (milliseconds to seconds) light curves. PDSs of BHXRBs often show one or multiple peak-like features called quasi-periodic oscillations (QPOs), along with strong broadband flat-top or red noise \citep{Psaltis99, Nowak00, D08, Belloni10, Ingram19}. Specifically, LFQPOs are detected across a frequency range from $\sim0.1$\,Hz to $\sim$30\,Hz \citep[e.g.][]{Belloni02, Casella04}. Based on their centroid frequency ($\nu$), quality factor $Q=\nu/\Delta\nu$ (where $\Delta\nu$ is the full width at half maximum, FWHM), fractional rms variability, broad-band noise components, and phase-lag behaviour, LFQPOs are classified as type A, B, and C 
\citep{Wijnands99, Homan01, Remillard02, Casella05}.
Models proposed to explain the origin of these LFQPOs focus
on mainly two mechanisms:
accretion-flow instabilities (e.g. magneto-acoustic waves: \citealt{Titarchuk98, Cabanac10}; spiral density waves: \citealt{Varniere02, Varniere12}) or changes in the geometry of the inner flow (e.g. Lense-Thirring precession: \citealt{Stella98, Stella99, Schnittman06, Ingram11}; relativistic dynamic frequency: \citealt{Misra20}; shock oscillation: \citealt{C93, MSC96, C15}).

Energy-dependent timing properties of LFQPOs, including fractional rms amplitude and time/phase lags, provide insight into the radiative processes responsible for QPOs. Studies have shown that LFQPOs are primarily generated by Comptonized photons \citep{Rao00, C05}, and the large fractional rms of type-C QPOs above $\sim20-30$\,keV, extending up to $\sim200$\,keV in \textit{Insight}-HXMT observations, suggests that the QPO mechanism is not directly associated with the thermal disk \citep{Casella04, Zhang17, Zhang20, Ma21}.

The time (or phase) lag between different X-ray energy bands can be quantified from the cross-spectrum, which measures the relative delay between variations in two bands. Conventionally, a positive (hard) lag indicates that hard photons lag behind soft photons, whereas a negative (soft) lag indicates that the soft photons arrive later than the hard photons. Both soft and hard lags have been reported for type-C QPOs in BHXRBs \citep{Cui00, Munoz-Darias10, Uttley11, Pahari13}. Based on an analysis of 15 BHXRBs, \citet{van17} found that soft lags are detected preferentially in high-inclination systems. In GRS~1915+105, \citet{Qu10} found an anticorrelation between QPO frequency and phase lag, with a change from hard to soft lag near $\sim2$\,Hz; a similar sign change was reported for XTE~J1550$-$564 around $\sim3.2$\,Hz \citep{Dutta16}.

The propagating-perturbation model \citep{Bottcher99, Lin00} relates lag sign changes to a change in the direction of perturbation propagation near a crossover frequency, although the physical origin of such a reversal remains uncertain. Within the context of the shock-oscillation framework, \citet{Dutta16} argued that multiple effects, including Comptonization, disk reflection, and gravitational lensing, can contribute to the observed lags, with their relative importance varying with inclination and source geometry. Soft lags can also coincide with enhanced radio emission in high-inclination systems, suggesting that outflows/jet activity may additionally influence the lag behaviour \citep{Patra19, AChatterjee20}.

Among the various transient BHXRBs discovered to date, Swift J1727.8-1613 is a newly identified black hole candidate that was discovered on 
2023 August 24 by Swift/BAT \citep{Kennea23}. Initially detected as GRB 230824A \citep{Page23}, MAXI/GSC detected a rapid increase in the source 
X-ray flux reaching upto $\sim 7$ Crab in the $2-20$ keV energy band \citep{Nakajima23, Negoro23}. It was recognized as a Galactic transient BHXRB 
following subsequent multi-wavelength observations (X-rays: \citealt{OConnor23, Sunyaev23, Dovciak23, Katoch23}; Optical: 
\citealt{Castro-Tirado23, Wang23, Sanchez23}; Near-infrared: \citealt{Baglio23}; Radio: \citealt{Miller-Jones23, Patra23, Trushkin23}). 
The outburst persisted for nearly six months, offering an unprecedented opportunity to study the accretion dynamics of a previously unknown 
BHXRB system. VLBI radio imaging revealed a north-south aligned, spatially resolved continuous jet during the early stages of the outburst 
\citep{Wood24}, while optical spectroscopic monitoring estimated an orbital period of approximately $7.6$ hours and estimated the distance of 
the source to be $2.7\pm0.3$ kpc, implying a significant elevation above the Galactic plane \citep{Sanchez24, Sanchez25}. 
Further optical data revealed complex inflow and outflow signatures, likely enhanced due to its proximity.

X-ray timing studies detected prominent type-C LFQPOs using multiple observatories including 
Swift/XRT, NICER, INTEGRAL, AstroSat, and \textit{Insight}-HXMT \citep{Palmer23, Draghis23, Katoch23, Yu24, Nandi24, Mereminskiy24, D25}. 
The IXPE mission made measurements of X-ray polarization of the source, suggesting that the Comptonizing corona is elongated along the 
disk plane rather than being spherical \citep{Veledina23, Ingram24}. The polarization angle was found to be perpendicular to the jet direction, 
hinting at a jet-corona connection.

Advanced phase-resolved polarimetric analysis, particularly during QPO cycles, showed modulation in the photon index but not in the polarization 
parameters, challenging existing models such as Lense-Thirring precession and calling for deeper theoretical investigations \citep{Zhao24}.
Spectral modeling using data from NICER, NuSTAR, and \textit{Insight}-HXMT during the hard-intermediate state suggested the presence of a rapidly 
spinning black hole (spin $\sim0.98$) and an inclination angle in the range of $30^\circ-50^\circ$ \citep{Peng24}. However, 
\citet[][hereafter Paper-I]{D24} estimated inclination angle of the source as $i \sim 85^\circ$, which aligns with the detection of soft time-lags.
Subsequent reflection-based analyses of the \textit{Insight}-HXMT data indicated that the optically thick disk extends to small radii during the intermediate state, with inner disk radii consistent with a disk close to the ISCO \citep{Cao25,Peng25}.

\citet[][hereafter Paper-II]{D25} using \textit{Insight}-HXMT data of the initial rising phase of the outburst, studied monotonic evolution of the type-C 
QPOs (from $0.09$ to $8.78$~Hz) with the propagating oscillatory shock (POS) model \citep{C08}. This is the time varying form the shock oscillation 
model \citep[SOM; see][]{MSC96, RCM97, C15}. According to the SOM model, strong type-C QPOs are originated due to resoance oscillation of the shock, 
while type-B QPOs occur either due to the non-satisfaction of the Rankine-Hugoniot condition or due to a weakly resonating Compton cloud, and broader 
type-A QPOs are attributed to weak oscillations of the shockless centrifugal barrier. In Paper-II, POS model fit of the evolving QPOs provided 
instantaneous location and strengths of the oscillating shock. The mass of the BHC Swift~J1727.8-1613 is also estimated as $13.5\pm1.9~M_\odot$ from 
spectro-temporal studies. 

Swift J1727.8-1613 has been the subject of several recent spectral and timing investigations during its 2023 outburst that characterized the evolution of QPO properties and energy-dependent time-lags \citep{Rawat25,Liao25}, including broadband studies using Insight-HXMT \citep{PeiJin25}. However, the connection between the broadband QPO-associated time-lags and a physically motivated dynamical scale of the Comptonizing region has not yet been explored.
In this work, we investigate the evolution of temporal properties such as QPOs and their associated time-lags in Swift J1727.8-1613, and examine their implications for accretion dynamics in black hole X-ray binaries. Utilizing the dense \textit{Insight}-HXMT monitoring of the rising phase, we track the lag evolution simultaneously across the $2-150$\,keV energy range and study the energy dependence of QPO frequency, fractional rms, and time-lags. We quantitatively relate the observed changes in lag sign and magnitude to the evolution of QPO frequency, spectral softness, and the shock location ($X_s$) inferred within the propagating oscillatory shock (POS) framework (Paper~II). This joint spectro-temporal approach allows us to identify energy-dependent zero-crossing points in the lag evolution and to place the observed lag behavior in the context of a shrinking and expanding Comptonizing region. The structure of this paper is as follows: \S2 describes 
the observational data and reduction methods used in our analysis. \S3 presents the key results, including the evolution of time-lag properties 
and their energy-dependent behavior. In \S4, we discuss the implications of these findings within the framework of current QPO models, highlighting 
both consistencies and discrepancies. 
Finally, \S5 summarizes our conclusions.

\vskip 1.0cm

\section{Observation and Data Analysis}

In this section we describe the \textit{Insight}-HXMT observations used in this work and summarize the data reduction and timing-analysis procedures employed to compute the PDS and lag spectra.

\subsection{Data Reduction}

We study archival data of 44 observations of \textit{Insight}-HXMT \citep{Zhang14, ZhangSN20} from 2023 August 25 (MJD=60181.42) to 2023 October 05 (MJD=60222.20). 
\textit{Insight}-HXMT carries three scientific payloads: 
the LE telescope, consisting of three SCD detectors with an effective area of 
$384$\,cm$^{2}$ operating in $1-15$\,keV \citep{Chen2020};
the ME telescope, consisting of 1728 Si-PIN detectors with an effective area of $952$\,cm$^{2}$ operating in $5-30$\,keV \citep{Cao2020}; 
and the HE telescope, consisting of 18 NaI/CsI phoswich detectors, each with an effective area of $5,100$\,cm$^{2}$, operating in $20-250$\,keV \citep{Liu2020}.
Level-1 data were downloaded from the HXMT archive and processed to generate cleaned Level-2 products using \texttt{HXMTDAS}\footnote{\url{http://hxmten.ihep.ac.cn/software.jhtml}} v2.05 following standard procedures.
The \textit{hpipeline} task was used to automatically apply data screening and calibration for all three instruments.
Good time intervals were selected by excluding periods affected by Earth occultation, South Atlantic Anomaly (SAA) passages, unstable pointing, and low geomagnetic cutoff rigidity, adopting criteria of elevation angle $> 10^{\circ}$, cutoff rigidity $> 8$\,GeV, pointing offset $< 0.04^{\circ}$, and $> 600$\,s away from the SAA.
Source light curves for the HE, ME, and LE instruments were extracted using the standard \textit{helcgen}, \textit{melcgen}, and \textit{lelcgen} tasks respectively.
To study properties of QPOs, Time, and Phase Lags, we generated $0.01$\,s time binned light curves in four different energy bands: $2-4$\,keV of LE, 
$4-10$\,keV of LE, $10-30$\,keV of ME, and $30-150$\,keV of HE. 
To study energy dependent time-lag, 
we divided $2-150$~keV band data of three instruments into following energy bands: $2-4$, $4-10$, $10-15$, $15-20$, $20-25$, $25-30$, 
$30-35$, $35-40$, $40-45$, $45-50$, $50-60$, $60-70$, $70-80$, $80-90$, $90-100$, $100-150$\,keV and generated light curves in $0.01$\,s time bin.
To study the variation of the X-ray flux and hardness ratios, daily average MAXI/GSC light curves are obtained from the 
MAXI website \footnote{\url{https://maxi.riken.jp/star_data/J1727-162/J1727-162.html}} and Swift/BAT daily light curves are obtained from the 
Swift website \footnote{\url{https://swift.gsfc.nasa.gov/results/transients/weak/SWIFTJ1727.8-1613/}}.
The instrumental count rates in different energy bands are converted to physical flux in mCrab units with conversion factors mentioned in the respective websites \footnote{\url{https://maxi.riken.jp/top/readme.html}} \footnote{\url{https://swift.gsfc.nasa.gov/results/transients/BlackHoles.html}}.

\subsection{Data Analysis}

We perform temporal analysis using the \texttt{Stingray} software package \citep{Huppenkothen19,Bachetti24}. To construct power density spectra (PDS), we divide the $0.01$\,s time-binned light curve into $81.92$\,s segments, compute individual power spectra, and then average them. The Poisson noise level is estimated from the mean power in the $30$--$50$\,Hz frequency range and subtracted from the averaged PDS, after which the PDS is geometrically rebinned; this procedure is repeated for each energy band. We model each PDS using multiple Lorentzian components in {\fontfamily{qcr}\selectfont XSPEC} version 12.13.0c \citep{Arnaud96}, accounting for broadband noise, the QPO peak, and its harmonic. From the fits we extract the centroid frequency ($\nu_{0}$), width ($\Delta\nu$), quality factor ($Q=\nu_{0}/\Delta\nu$), and fractional rms amplitude, where the rms is computed by integrating the best-fitting Lorentzian component over frequency and taking the square root of the integrated power.

Phase-lag and time-lag spectra are computed using the complex cross-spectrum between a given energy band and a reference band, following standard Fourier techniques implemented in \texttt{Stingray}. The phase lag is obtained from the argument of the averaged cross-spectrum, and the corresponding time lag is calculated by dividing the phase lag by $2\pi\nu$. 
To determine the characteristic lag associated with individual variability components, we compute the average lag over the full width at half maximum ($\Delta\nu$) of the corresponding Lorentzian component, centered around the 
centroid frequency ($\nu_{0}\pm\frac{\Delta\nu}{2}$; \citealt{Reig00}).
Lag-energy spectra are produced by repeating this procedure for multiple energy channels with respect to the same reference band.

\section{Results}

In this section we present the observational results, starting with the long-term outburst evolution and hardness ratios, followed by the evolution of the QPO properties and time-lags, and finally the correlations and energy-dependent lag behaviour.

\subsection{Evolution of Outburst Profile and Hardness Ratio}

\begin{figure*}
\vskip 0.5cm
  \centering
    \includegraphics[angle=0,width=16cm,keepaspectratio=true]{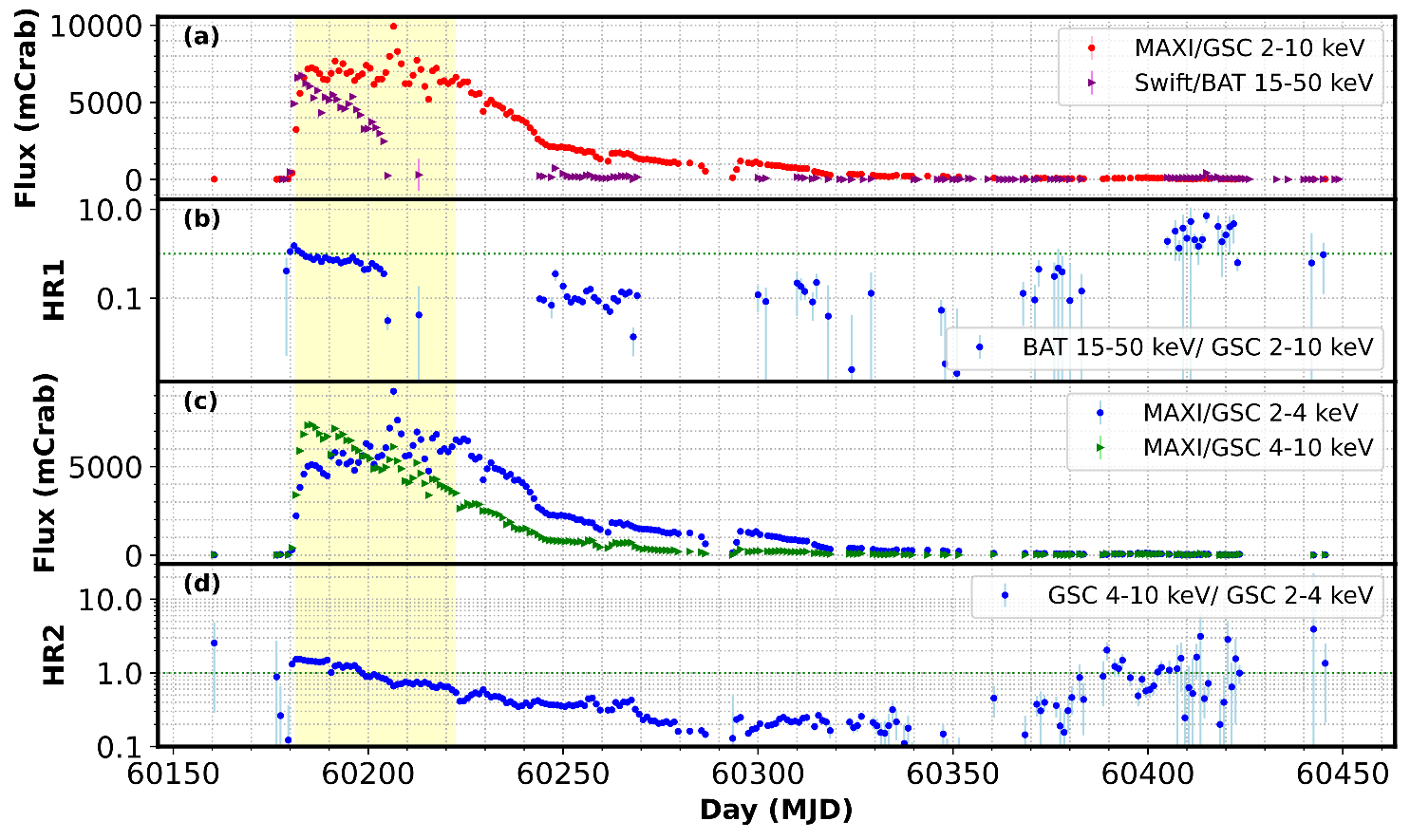} 
	\caption{Evolutions of (a) 2-10 keV MAXI/GSC flux and 15-50 keV Swift/BAT flux in units of mCrab 
	         (b) HR1, i.e., the ratio of 15-50 keV Swift/BAT flux to 2-10 keV MAXI/GSC flux 
	         (c) MAXI/GSC flux in 2-4 keV and 4-10 keV range in units of mCrab  
	         (d) HR2, i.e., the ratio of 4-10 keV to 2-4 keV MAXI/GSC flux are shown with time (MJD). 
		 The yellow shaded region indicates the time when LFQPOs are present in the HXMT lightcurves. 
         }
    \label{lc_hr}
\end{figure*}

Fig. \ref{lc_hr} shows the evolution of outburst X-ray fluxes in different energy bands and their hardness ratios from 2023 August 4 (MJD=60160) 
to 2024 May 15 (MJD=60445) in different panels. 
Panel (a) presents the variation in Swift/BAT ($15-50$\,keV) flux and MAXI/GSC ($2-10$\,keV) flux, 
while panel (b) illustrates the changes in their corresponding hardness ratio (HR1). Similarly, panel (c) displays the evolution of MAXI/GSC flux 
in the lower ($2-4$\,keV) and higher ($4-10$\,keV) energy bands, with panel (d) depicting the variation in their hardness ratio (HR2). 
From the figure, we can see that after the outburst started on 2023 August 24 (MJD=60180), both the soft and hard fluxes started to increase rapidly. 
Within $\sim4$ days from the start of the outburst, the Swift/BAT $15-50$\,keV flux attained a maximum on $\sim$ MJD 60184 and started to decrease 
gradually. The MAXI/GSC $4-10$\,keV flux also showed a similar behavior, attaining the maximum value on $\sim$ MJD 60185 and decreasing afterwards.
The MAXI/GSC $2-10$\,keV and $4-10$\,keV flux attained an initial maximum on $\sim$ MJD 60185, showed multiple reflares till MJD 60231, and stared 
to decrease gradually. Looking at the hardness ratios in panel (b) and panel (c), we can see that at the start of the outburst, both HR1 and HR2 had 
a value of $\sim1.5$, which indicates an initial hard state. Both of the hardness ratios started to decrease gradually after that.
After $\sim$ MJD 60190, the hardness ratios became smaller than unity, which indicates that the outburst moved into the soft state.
They continued decreasing upto $\sim$ MJD 60300. For the next $\sim$ 50 days, the HR1 showed random variation, while the HR2 remained stationary 
around $\sim 0.2$, indicating a clear dominance of the soft band photons. After MJD 60350, the hardness ratios started showing an increasing trend, 
which indicates the outburst moved into the declining harder spectral states.

\subsection{Evolution of Quasiperiodic Oscillations (QPO) and Time Lag}

\begin{figure}
\vskip 0.1cm
  \centering
    \includegraphics[angle=0,width=9.0cm,keepaspectratio=true]{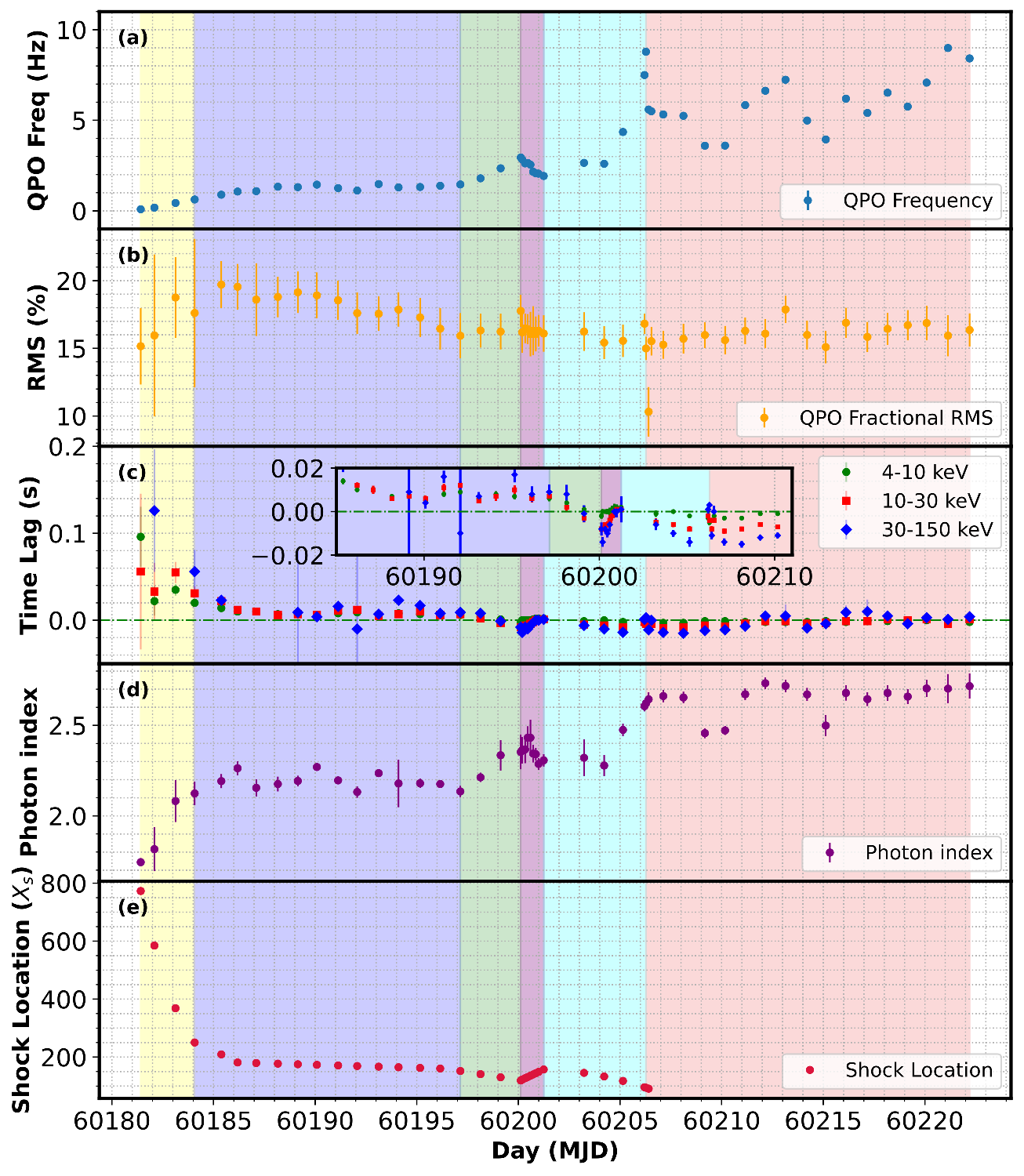} 
	\caption{Evolutions of (a) LFQPO centroid frequency in \textit{Insight}-HXMT 10-30 keV energy band lightcurves,
                            (b) Fractional RMS of the LFQPO,
                        (c) QPO time-lag of 4-10 keV, 10-30 keV and 30-150 keV enrgy bands with respect to 2-4 keV band,
                        (d) Photon index obtained from the spectral fit with the {\fontfamily{qcr}\selectfont pexrav} model,
                        (e) Shock location from the {\fontfamily{qcr}\selectfont POS} model fit with time.
                        Different color highlights indicate the six different phases of evolution of LFQPO frequency as mentioned in Paper-II.}
 \label{lag-evo}
\end{figure}

Strong type-C QPOs are detected during the rising hard-intermediate phase of the outburst (yellow highlighted period in Fig.~\ref{lc_hr}) in the PDS of all three \textit{Insight}-HXMT instruments (LE: $2$--$10$\,keV, ME: $10$--$30$\,keV, and HE: $30$--$150$\,keV). Figure~\ref{lag-evo} summarizes the evolution of the QPO centroid frequency, QPO fractional rms, and the QPO time-lags of the $4$--$10$\,keV, $10$--$30$\,keV, and $30$--$150$\,keV bands with respect to the $2$--$4$\,keV band, together with the photon index from the combined (LE+ME+HE) spectral fits and the shock location from the POS model (both adopted from Paper-II).

We also fitted the PDS in the LE and HE bands and found that the QPO frequencies are broadly consistent with those measured in the ME band; details of the QPO evolution are presented in Paper-II. Following Paper-II, we divide the rising phase into six intervals (shaded in Fig.~\ref{lag-evo}): (1) MJD 60181.42--60184.07 (yellow), (2) MJD 60184.07--60197.15 (blue), (3) MJD 60197.15--60200.12 (green), (4) MJD 60200.12--60201.25 (purple), (5) MJD 60201.25--60206.28 (cyan), and (6) MJD 60206.28--60222.20 (pink).
Looking at the panel (a) of Fig. \ref{lag-evo}, we see that the QPOs appear on the very first day of the HXMT observtion, i.e. on MJD 60181.
The QPO frequency was $0.088\pm0.003$\,Hz on this starting day. After that, during the first phase the QPO frequency started to increase 
steadily upto MJD 60185 to a value of $0.890\pm0.004$\,Hz, and became more or less steady  for $\sim12$ days in the second phase. 
Starting from a value of $\sim15\%$, the fractional RMS values increased upto $\sim20\%$ in the first phase of evolution (till MJD 60185) and
decreased to about the starting level in the second phase of evolution (till MJD 60197). The time-lags were positive in all three bands 
initially, with a value of $0.10\pm0.03$\,s in $4-10$\,keV, and $0.06\pm0.09$\,s in $10-30$\,keV bands on MJD 60181 and $0.13\pm0.07$\,s 
in $30-150$\,keV band on MJD 60182. All the lags started to decrease rapidly during the first phase and became steady around $\sim0.01$\,s 
after $\sim$ MJD 60188. The photon index was $1.74\pm0.02$ initially, indicating a hard spectral state. It then started to increase rapidly 
as the spectra became softer in the first phase, and reached a value of $\sim 2.2$ around MJD 60185 and stayed at a roughly constant level 
in the second phase of evolution. 
The shock was away at $\sim 773\,r_s$ ($r_s$: Schwarzschild radius) from the BH initially, and rapidly started to move inward during 
the first phase. After MJD 60186, the shock stayed roughly at a constant location of $\sim 169\,r_s$ in the second phase till $\sim$ MJD 60197.

After the second phase, all of the above quantities started to show more variation within shorter time periods. The QPO frequency increased 
and decreased rapidly multiple times, showing some local maximum peaks. These local maximum peaks were coincident with the reflares in the 
soft energy band. In the third phase, i.e. after MJD 60197, both the QPO frequency and photon index increased while the QPO fractional RMS 
was roughly steady at $\sim16\%$. The time-lags in all three energy bands started to decrease in this phase and switched sign around 
$\sim$ MJD 60199 and became more negative. The shock location also decreased and reached a value of $\sim119\,r_s$ on MJD 60200.
However, in the fourth phase, the shock started to move outward rapidly and reached $\sim157\,r_s$ in just one day. The time-lags started 
to become less negative and became positive in all the three energy bands during this period. The QPO frequency decreased in this phase, 
while both the QPO fractional RMS and photon index stayed roughly constant. After MJD 60201, in the fifth phase, the shock again started 
to move inward and reached $\sim91\,r_s$, the time-lags again became negative while the QPO frequency started to increase.
The QPO fractional RMS was roughly constant. However, the photon index started to increase in this phase which indicated an increase 
in the soft X-ray emission. In the fifth phase, the photon index was roughly constant at $\sim 2.7$. The QPO fractional RMS was also 
remained roughly constant at $\sim16\%$. The time-lags in $4-10$\,keV and $10-30$\,keV bands were negative in this phase, while the 
time-lag in $30-150$\,keV band showed negative and positive values in a narrow range around $0$\,s.

\begin{figure*}
\vskip -0.2cm
  \centering
    \includegraphics[angle=0,width=16cm,keepaspectratio=true]{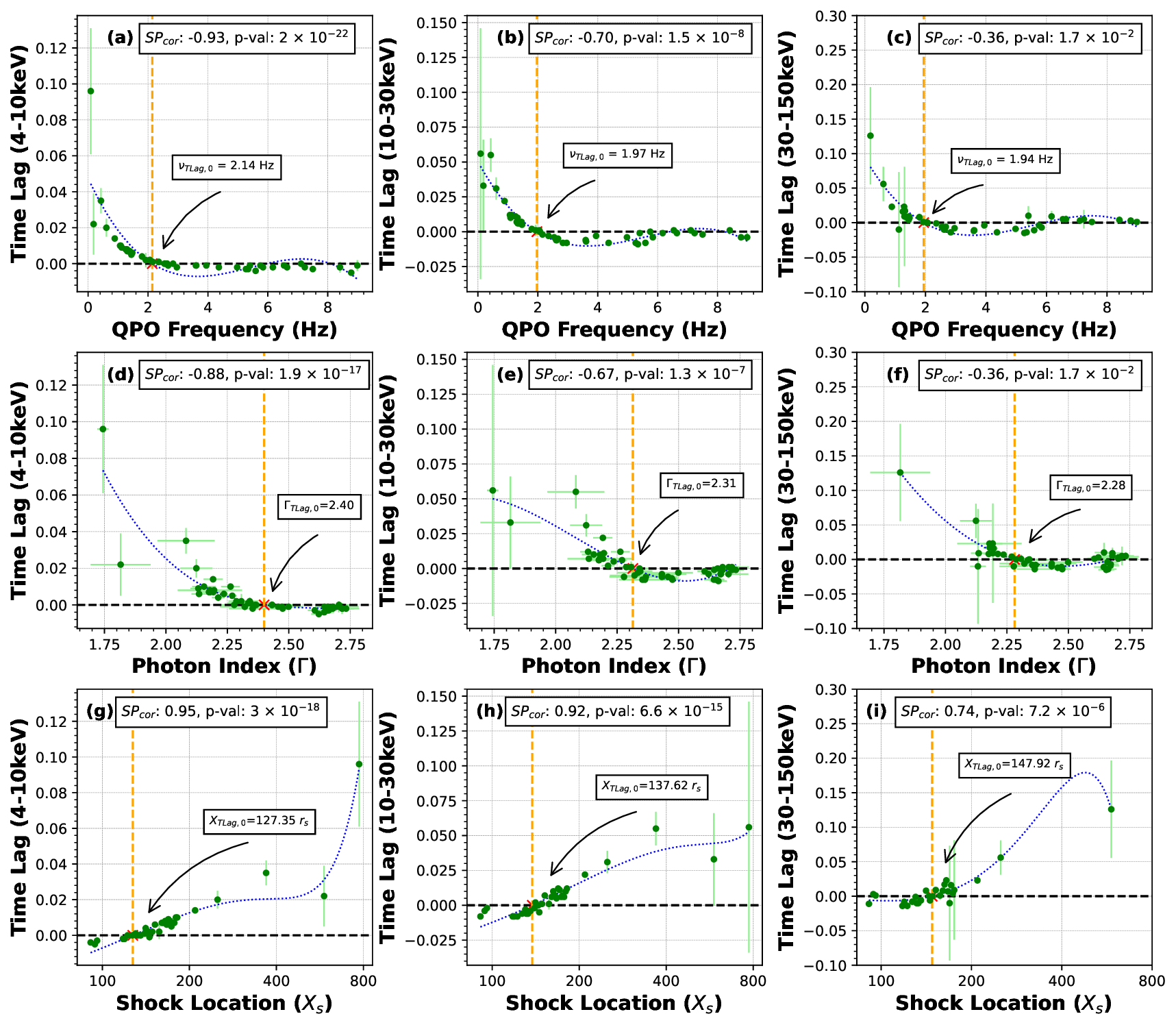} 
	\caption{Variation of QPO time-lags (TLag) in the LE ($4$-$10$~keV), ME ($10$-$30$~keV), and HE ($30$-$150$~keV) bands (with respect 
	to the $2$-$4$~keV band) with QPO frequency, {\fontfamily{qcr}\selectfont pexrav} model fitted photon index ($\Gamma$), POS model fitted shock location ($X_s$) is shown. 
    Spearman rank correlation coefficients and their 
	p-values are mentioned insets.
    Vertical dashed orange line marks the point where the time lag crosses zero, i.e. transitions from hard lag to soft lag.
    The zero-crossing frequencies ($\nu_{TLag,0}$), zero-crossing photon indices ($\Gamma_{TLag,0}$) and zero-crossing shock locations ($X_{TLag,0}$) have been mentioned in the plots. 
    The blue dotted lines are the cubic spline functions which are employed here to determine the zero-crossing point.
    }
\label{corr}
\end{figure*}

\subsection{Correlations of Time-lags with QPO Frequency, Photon Index, and Shock Location}

To understand the nature of the variation of TLag in more detail in different energy bands of HXMT/LE, ME, and HE 
with QPO frequency ($\nu_{QPO}$), {\fontfamily{qcr}\selectfont pexrav} model-fitted photon index ($\Gamma$), and {\fontfamily{qcr}\selectfont POS} 
model-fitted shock location ($X_s$), we studied the correlations among them. The $\Gamma$ and $X_s$ values are adopted from Paper-II. 
To quantify the strength and direction of the monotonic relationship between TLag and each parameter, we compute the Spearman rank correlation coefficient ($SP_{cor}$) and its associated p-value. These quantities are reported in the insets of Fig.~\ref{corr}.

The time lags in our data show a sign change, transitioning from positive to negative, indicating a potential crossover in the physical process responsible for the observed delays. 
Identifying the precise point at which the lag crosses zero is important for understanding the evolution of the accretion flows and any underlying changes in geometry or emission mechanisms.
Initially, we attempted to determine the zero-crossing points by fitting the lag variations using simple polynomial models. 
However, these fits were not statistically acceptable - likely due to the complex nature of the lag variation.
To overcome this, we adopted a cubic spline interpolation method, which provides a more reliable approach for modeling continuous functions through a set of discrete data points.
By applying cubic spline interpolation to the lag data, we were able to accurately trace the behavior of the lag curve and identify the zero-crossing points.

\subsubsection{Correlation between QPO Frequency and Lags}

We investigated the relationship between the QPO frequencies ($\nu_{\text{QPO}}$) and the time lags (TLag) in different hard X-ray bands of 
{\it Insight}-HXMT, namely LE ($4$--$10$\,keV), ME ($10$--$30$\,keV), and HE ($10$--$30$\,keV), measured with respect to the soft X-ray band ($2$--$4$\,keV), as shown in (Fig.~\ref{corr}a-c). 
To get a quantitative estimate of the strength and direction of these correlations, we used the Spearman rank correlation method.
The computed $SP_{cor}$ and their associated p-values are indicated within each panel of Fig.~\ref{corr}.
The results reveal a consistent negative (anti-) correlation between TLag and $\nu_{\text{QPO}}$ across all the three energy bands.
The LE and ME bands show strong anti-correlations, as evident from the large negative $SP_{cor}$ values ($\le-0.7$) and extremely small p-values, implying that as the QPO frequency increases the observed time lags decrease, which suggests a causal connection between the time lags and geometry of the Comptonizing region.
In contrast, the HE band exhibits a weaker anti-correlation, with a more modest $SP_{cor}$ value of $-0.36$ and a higher p-value ($0.017$), which still suggests a statistically significant trend but with lower confidence and strength compared to the lower energy bands.
This decrease of correlation strength at higher energies may indicate that the time lag behavior becomes more complex or less coherent in this larger energy band ($30-150$\,keV) potentially due to increased contributions from higher energy photons having different emission regions or scattering histories.

To further explore the $\nu_{\text{QPO}}$-TLag relationship, we determine the zero-crossing QPO frequency ($\nu_{TLag,0}$) at which the time lag transitions from positive (hard lag) to negative (soft lag) using cubic spline interpolation of the $\nu_{\text{QPO}}$-TLag data.
The resulting values of $\nu_{TLag,0}$ are $2.14$\,Hz, $1.97$\,Hz and $1.94$\,Hz in the LE, ME and HE bands respectively.
The gradual shift of $\nu_{TLag,0}$ to lower QPO frequencies with increasing energy suggests that the transition from hard to soft lags occurs under different physical conditions for photons of different energies.

\subsubsection{Correlation between Photon Index and Lags}

The correlations between the time-lags (TLag) and {\fontfamily{qcr}\selectfont pexrav} model-fitted photon indices ($\Gamma$) are shown in Fig.~\ref{corr}(d-f). 
The $\Gamma$ values are adopted from the broadband \textit{Insight}-HXMT spectral fits ($2$-$150$~keV) reported in Paper-II. 
Similar to the $\nu_{\text{QPO}}$-TLag correlations, the statistical significance of the $\Gamma$-TLag correlations 
in the three hard X-ray bands of HXMT is studied using the Spearman rank correlation method.
The values of $SP_{cor}$ and the p-values have been mentioned inside the respective figures. 
The values indicate the presence of consistent negative (i.e., anti-) correlations between the photon index and time-lags.
In particular, the LE and ME bands show strong negative correlations, suggesting that as the spectrum softens (i.e., $\Gamma$ increases), the observed time lags decrease.
On the other hand, the HE band displays a weaker correlation, indicating that the relationship between $\Gamma$ and TLag becomes less pronounced at higher energies, possibly due to the large ($30-150$\,keV) energy interval where photons with different energies have different emission regions having competing contributions to the net time lag.

To further probe the evolution of the lag behavior with spectral changes, we estimate the zero-crossing photon indices ($\Gamma_{TLag,0}$) where the time-lags transition from positive to negative values using cubic spline interpolation of the $\Gamma$-TLag data.
The $\Gamma_{TLag,0}$ values are obtained as $2.40$, $2.31$ and $2.28$ in the LE, ME and HE bands respectively.
These values indicate that the lag sign reversal occurs at progressively lower photon indices as energy increases, suggesting that the transition to soft lags happens earlier in the spectral evolution for higher-energy photons.

\begin{figure}
\vskip 0.5cm
  \centering
    \includegraphics[angle=0,width=9.0cm,keepaspectratio=true]{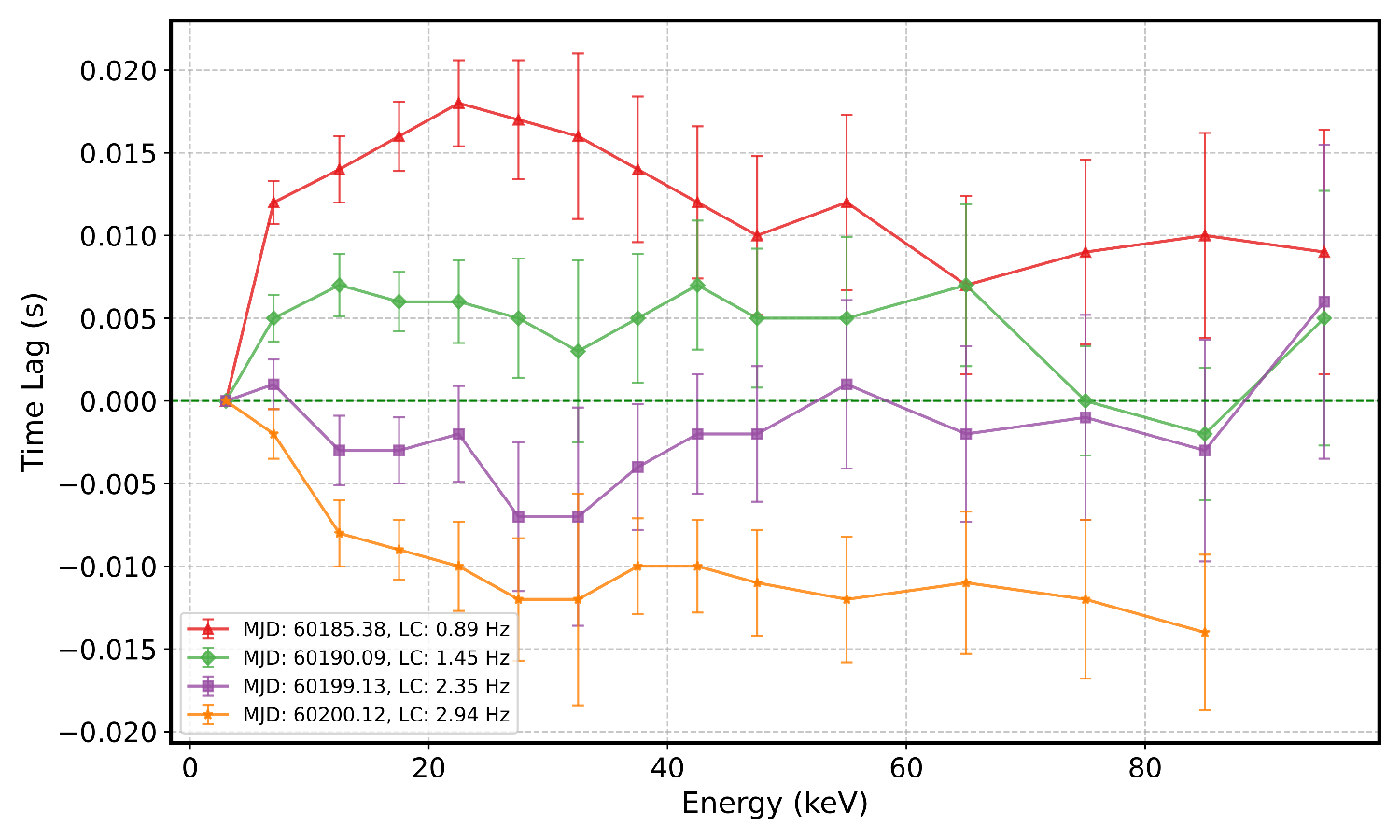} 
	\caption{Energy dependence of QPO time-lag (w.r.t. 2-4 keV) for different QPO frequencies. Here the label `LC' refers to the centroid frequency of the best fit Lorentzian component to the QPO peak. (centroid frequency of the QPO).}
    \label{eng-dep}
\end{figure}

\subsubsection{Correlation between Shock Location and Lags}

The correlation between the shock locations ($X_s$) and time-lags (TLag) is shown in Fig.~\ref{corr}(g-i). 
The $X_s$ values are adopted from the {\fontfamily{qcr}\selectfont POS} model fit of type-C QPO evolutions of Paper-II. 
Similar to the previously mentioned correlations, we studied the statistical significance of the $X_s$-TLag correlations in the three hard X-ray bands of HXMT using the Spearman rank correlation method.
The estimated $SP_{cor}$ values and their associated p-values, displayed within each panel of Fig.~\ref{corr}, reveal strong positive correlations between shock location and time lag across all energy bands. 
The correlation is particularly strong in the LE and ME bands, where the $SP_{cor}$ values are very high ($>0.9$) and the p-values are extremely low, indicating that these correlations are statistically significant. 
In this case, even in the HE band, where $SP_{cor}$ is relatively lower ($0.74$), the p-value still confirms the presence of a significant trend.
These results suggest that as the shock moves outwards, i.e., the size of the Comptonizing region or post-shock corona increases, the observed time lag between hard and soft photons also increases.

To gain further insight into this relationship, we determine the zero-crossing shock location ($X_{TLag,0}$) at which the time lag transitions from positive to negative using cubic spline interpolation of the $X_s$-TLag data.
The resulting $X_{TLag,0}$ values are obtained to be $127.35\,r_s$, $137.62\,r_s$, and $147.92\,r_s$ respectively in LE, ME and HE bands.
These results imply that as the size of the post-shock Comptonizing corona decreases, the energy of the Comptonized photons decreases and the mechanisms that produce soft lag becomes dominant.

\subsection{Evolution of the Energy Dependent Time-Lags}

We show the energy dependent variation of the QPO time-lags for four different MJDs with different QPO frequencies in Fig.~\ref{eng-dep}.
The time-lags of $4-10$, $10-15$, $15-20$, $20-25$, $25-30$, $30-35$, $35-40$, $40-45$, $45-50$, $50-60$, $60-70$, $70-80$, $80-90$, $90-100$\,keV
bands are calculated with respect to the $2-4$\,keV band. 
On MJD 60185, when the QPO frequency was relatively low at $0.89$\,Hz, the time-lags were consistently positive across all energy bands. 
The lags increased with energy up to the $20-25$\,keV band, reaching a maximum value of $\sim18$\,ms, and then gradually decreased at higher energies, settling around $9$\,ms in the $90-100$\,keV band.
On MJD 60190, as the QPO frequency increases to $1.45$\,Hz, a reduction in the magnitude of time-lags was observed across all energy bands.
The time-lags increase upto $10-15$\,keV, and stay almost constant upto $90-100$\,keV, except for zero lag in $70-80$\,keV 
and negative lag in $80-90$\,keV bands. 
Subsequently, on MJD 60199, the QPO frequency increases to $2.35$\,Hz, and time-lags for most of the 
energy bands become negative, except for $10-15$\,keV and $50-60$\,keV bands. 
Finally, on MJD 60200, when the QPO frequency reached $2.94$\,Hz, the time-lags were entirely negative across all energy bands.

\section{Discussions}

In this work, we investigate outburst profile of the source Swift J1727.8-1613 during its $2023-2024$ discovery outburst that 
lasted for $\sim10$\,months. We study the temporal properties, specifically the evolution of QPOs and associated time-lags, 
during the initial rising phase of the outburst, from 2023 August 25 (MJD 60181) to 2023 October 05 (MJD 60222).
For our temporal analysis, we utilize the $0.01\,s$ time-binned lightcurves from the LE, ME and HE instrument of {\it Insight}-HXMT satellite.
We construct the cross-spectrum of different energy bands as mentioned in \S2 and estimate the time-lag with respect to the lowest energy band.
Our analysis reveals significant evolution in the time-lags between different X-ray energy bands, correlated with changes in the QPO frequency, 
spectral state, and the size of the Comptonizing region.

The most striking result is the observed transition from positive (hard) time-lags to negative (soft) time-lags as the outburst evolved.
A similar hard-to-soft lag transition in Swift J1727.8-1613 was independently reported using NICER observations (see Fig.~5 of \citealt{Rawat25}), lending strong support to the robustness of this behavior across different instruments and energy ranges.
Initially 
(MJD $60181-60197$), when the source was in a harder state (lower photon index $\Gamma \sim 1.7 - 2.2$) and exhibited lower frequency type-C QPOs 
($\nu_{\mathrm{QPO}} \lesssim 1.4~\mathrm{Hz}$), time-lag in all the three harder energy bands ($4-10$\,keV, $10-30$\,keV, $30-150$\,keV) was positive, 
i.e., these hard photons consistently lagged behind the soft band ($2-4$\,keV) photons.
The magnitude of these hard lags reached up to $\sim1.3$\,s (Fig.~\ref{lag-evo}c), particularly in the earliest observations.
This behavior is commonly attributed to Comptonization delays \citep{Miyamoto88, Nowak00, Reig00},
where soft seed photons gain energy through multiple inverse Compton scatterings in a hot corona. Photons scattered to higher energies typically 
undergo more scatterings, resulting in a longer path and a delayed arrival time relative to lower-energy photons. 
The observed energy dependence during this phase (Fig.~\ref{eng-dep}, MJD 60185), where the positive lag generally increases with energy up to 
$\sim25$\,keV, is consistent with this interpretation. Furthermore, the strong positive correlation found between the time-lag magnitude and 
the shock location ($X_s$) (Fig.~\ref{corr}g-i) supports this picture: a more distant shock ($X_s$ up to $\sim773~r_s$ initially) indicates 
a larger corona which naturally leads to larger Comptonization delays. However, as we can see from Fig.~\ref{eng-dep}, on MJD 60185, the positive lag 
values deviated from the general increasing nature after $\sim25$\,keV. Even within the corona, the temperature and density distributions are not 
uniform, and photons scattering from different parts of the corona might have different delays and energies, potentially causing non-monotonic 
energy dependence of lag \citep{Chatterjee17}. Moreover, photons that originate from regions closer to the black hole can be subjected to the 
gravitational focusing effect and undergo more delay on top of that due to Comptonization \citep{Dutta16}.

As the outburst progressed, the shock front moved inwards, coinciding with an increase in QPO frequency above $\sim1.5-2$\,Hz and further 
spectral softening ($\Gamma > 2.2$) around MJD 60199. During this phase, the time-lags decreased significantly, crossing zero and becoming 
predominantly negative (soft lags) in the $4-10$\,keV and $10-30$\,keV bands (Fig.~\ref{lag-evo}c). Moreover, in the fourth 
phase, i.e., after MJD 60200.12, the shock front moved rapidly outward within one day, and the negative value of the time-lags also decreased 
rapidly and became eventually positive on MJD 60201.25. This shows that there exists a strong positive correlation between the magnitude of 
time-lags and the shock location. Soft lags indicate that the soft band photons arrive later than the harder band photons. 
While Comptonization still operates, the emergence of soft lags suggests that other mechanisms become dominant or significantly modify 
the net lag. \citet{van17} systematically found soft lags predominantly in high-inclination sources. 
Hard photons from the corona irradiate the cooler Keplerian disk, which reprocesses them producing a softer reflected spectrum.
The reflected photons are delayed because they travel an extra path from the corona to the disk and then to the observer.
This delay can cause soft lags, particularly in high-inclination systems where the reflected component is more prominent along the line 
of sight \citep{Dutta16, Chatterjee17}. The high inclination angle ($i \sim 85^\circ$) estimated for the source in Paper-I makes it a prime 
candidate for such geometrical effects to be significant. On the other hand, hard photons from the Comptonization region can travel through 
outflowing material, such as a Return OutFlow (ROF) region \citep{AChatterjee19}, or the base of a jet \citep{Patra19} becoming softer through 
Compton downscattering. These downscattered soft photons (which might also experience gravitational focusing effects) are significantly delayed 
compared to the directly observed hard photons. These effect can contribute to the observed soft lags, given the detection of strong jet 
activity in Swift J1727.8-1613 \citep{Wood24}. This mechanism is also considered particularly relevant for high-inclination sources where 
such outflows might obscure or interact with the line of sight.

The observed strong anti-correlations between the time-lag and both the QPO centroid frequency ($\nu_{\mathrm{QPO}}$; Fig.~\ref{corr}a-c) and 
the photon index ($\Gamma$; Fig.~\ref{corr}d-f) indicate a systematic evolution in the accretion flow geometry as the outburst proceeds. 
Within the framework of the POS model, this change is understood as the gradual inward movement of the shock front, which marks the boundary of 
the Comptonizing corona. As this shock front propagates inward (i.e., decreasing shock location $X_s$), the Comptonizing region itself shrinks.
We emphasize that the shock location inferred from POS modeling represents a dynamical scale associated with the oscillating Comptonizing region. 
The characteristic scale of the Comptonizing region is determined by the dynamics of sub-Keplerian inflow and need not coincide with the inner edge of the optically thick disk \citep{C97,CD04}.
Independent spectral analyses of the same \textit{Insight}-HXMT observations using relativistic reflection modeling have shown that the optically thick disk extends to much smaller radii during the intermediate state (e.g., \citealt{Cao25,Peng25}). Therefore, a comparatively large characteristic radius inferred from variability studies does not necessarily imply a strongly truncated disk, but rather reflects the spatial extent and dynamical state of the hot sub-Keplerian flow forming the Comptonizing region.
This reduction in size of the Comptonizing region leads to a decrease in the dynamical timescale of its oscillations, thereby causing an increase in the observed QPO 
frequency, which follows an inversely proportional relationship with shock location $\nu_{\mathrm{QPO}} \propto 1/{X_s(X_s-1)^{1/2}}$ 
\citep{C08, D10}. 
At the same time, the accretion flow undergoes spectral softening (increasing $\Gamma$). This is thought to be due to the enhanced cooling of the 
Comptonizing region by an increasing Keplerian disk accretion. In the hard state, or during the early stages of an outburst when the Comptonizing 
region is more extended and the shock is further away from the black hole, Comptonization is the primary mechanism producing hard lags. 
As the shock location decreases, the Comptonizing region contracts and the system transitions towards softer spectral states, the magnitude of this 
Comptonization-induced hard lag diminishes. In systems having high inclination, this decrease in the hard lag allows soft lags due to disk reflection, 
gravitational focusing and downscattering in outflows to become increasingly dominant, leading to a net reversal in the sign of the observed lag.

The frequency at which this reversal occurs, which we denote as $\nu_{\text{TLag,0}}$, obtained using spline interpolation of the lag-QPO frequency relation (Fig.~\ref{corr}a-c), shows variation 
with the energies. 
We find $\nu_{\text{TLag,0}} \sim 2.14$\,Hz in the LE band, $\sim1.97$\,Hz in the ME band, and $\sim1.94$\,Hz in the HE band.
The observed hard-to-soft lag transition as a function of $\nu_{\text{TLag,0}}$ is qualitatively similar to that seen in other well-studied BHXRBs like 
GRS 1915+105 (Qu et al. 2010, Dutta et al. 2018) and XTE J1550-564 (Dutta \& Chakrabarti 2016), although the specific transition frequencies differ 
between sources, potentially due to variations in black hole mass, spin, accretion rate, or specific corona/disk geometry.
The corresponding zero-crossing values of the shock location, denoted as $X_{TLag,0}$ have been found to be $127.35\,r_s$ in the LE band, $137.62\,r_s$ in the ME band, and $147.92\,r_s$ in the HE band.
This transition shock location values effectively mark the size of the shrinking Comptonizing region, below which it can no longer produce 
enough hard-lag, and the competing soft-lag mechanisms become predominant.
Moreover, the $\nu_{\text{TLag,0}}$ or $X_{\text{TLag,0}}$ values show energy-dependent trend, i.e., the hard-to-soft lag transitions occur at progressively lower QPO frequency or equivalently, at larger shock radii for photons of higher energy bands.
This implies that dominance of soft lags is more sensitive to the inner accretion flow configuration at higher energies.
High-energy photons originate from the innermost, hottest regions of the accretion flow, and are more susceptible to strong gravitational effects and interactions with outflows.
These soft-lag producing effects might become significant enough to cause the net lag to switch to a soft lag.
The $\Gamma_{TLag,0}$ values also show energy-dependent trend, i.e. the hard-to-soft lag transitions occur at progressively higher photon indices as energy of the photons decrease.
This implies the transition to negative lags happens earlier in the spectral evolution for higher-energy photons, as the energy of the Comptonized photons decrease in the later softer states.
The above results reinforce the idea 
that the lag evolution is a fundamental characteristic linked to changes in the geometry of the Comptonizing region during state transitions. 
The results provide observational backing for models like the POS, where the QPO frequency and the properties of the Comptonizing region 
(and thus the lags) are physically linked through the dynamics of an inward-propagating shock. The consistency between the lag evolution, 
the spectral softening, the increasing QPO frequency, and the decreasing shock location provides a consistent picture of the accretion flow 
dynamics during the rising phase of the outburst.

The shock compression ratio ($R$) listed in Table \ref{tab:table_qpo} represents the ratio of the post-shock to pre-shock matter density across the oscillating shock within the POS framework. 
In this framework, $R\approx4$ corresponds to a strong shock with maximum compression, whereas $R\to1$ indicates a progressively weakening shock as radiative cooling becomes more efficient. As the post-shock region acts as the Comptonizing region, $R$ directly influences the electron density and optical depth of this region. In our analysis, $R$ decreases from $\sim4$ in the early hard state to values close to unity as the outburst progresses, implying reduced compression and decreasing optical depth. This trend occurs simultaneously with inward shock motion and increasing QPO frequency, consistent with a contracting and progressively cooling Comptonizing region, as also inferred from independent Comptonization-based timing studies of this source.

We note that an alternative and complementary approach to constraining the properties of the Comptonizing medium has been developed through time-dependent Comptonization models, which utilize the combined evolution of rms amplitudes and time lags to infer coronal parameters such as size, temperature, and optical depth \citep[e.g.,][]{Karpouzas20,Bellavita22}. 
This framework has recently been applied to Swift J1727.8-1613 by \citet{Rawat25} and \citet{PeiJin25}, who interpreted the observed lag and variability properties in terms of changes in the physical conditions of the corona during the outburst, and reported a progressive reduction in the effective size of the Comptonizing region as the source evolved from harder to softer spectral states.
While these studies provide important constraints on the microphysical properties of the Comptonizing region, 
the present work focuses on a complementary, dynamical perspective, in which the evolution of time lags is directly linked to the propagation of an oscillatory shock and the associated changes in the geometry of the hot, sub-Keplerian flow. 
In this picture, the QPO frequency serves as a tracer of the characteristic dynamical scale of the Comptonizing region, allowing us to relate the observed lag reversal and its energy dependence to the inward motion and contraction of the corona.
The decreasing shock location inferred from the QPO frequency evolution naturally implies a contracting Comptonizing region, which is in agreement with the shrinking corona inferred by \citet{Rawat25} and \citet{PeiJin25}.
This broad consistency suggests that both approaches are probing the same underlying evolution of the accretion flow, albeit from different and complementary physical viewpoints.

\section{Conclusions}

In this section we briefly summarize the key observational results and their physical implications.

\begin{itemize}

  \item We report a clear evolution of time-lags during the rising phase of the outburst, including a transition from hard to soft lags as the source moves from harder to softer spectral states.

  \item The magnitude and sign of the time-lags are strongly correlated with the QPO centroid frequency, photon index, and the inferred size of the Comptonizing region, indicating a coupled evolution of timing and spectral properties.

  \item Hard lags at low QPO frequencies are consistent with Comptonization delays in an extended corona, whereas soft lags at higher QPO frequencies likely arise from disk reflection, gravitational focusing, and downscattering in outflows, particularly in this high-inclination system.

  \item The energy dependence of the lag reversal frequency suggests that higher-energy photons probe deeper and more dynamic regions of the accretion flow, making them more sensitive to geometric changes in the inner flow.

  \item Overall, our results demonstrate that time-lag evolution provides a powerful diagnostic of accretion flow geometry and its dynamical evolution during state transitions in black hole X-ray binaries.

\end{itemize}

\section*{Acknowledgements}

We are thankful to the anonymous referee for the insightful suggestions to improve the quality of the paper.
This work made use of archival data from the {\it Insight}-HXMT mission, a satellite project of China National Space Administration (CNSA) and Chinese Academy of Sciences (CAS).
S.K.N. acknowledges support from the visiting research grant of National Tsing Hua University.
D.D. acknowledge the visiting research grant of National Science and Technology Council (NSTC), Taiwan (NSTC 113-2811-M-007-010). 
H.-K. C. is supported by NSTC of Taiwan under grant NSTC 113-2112-M-007-020.

\section*{Data Availability}

We used archival data of \href{http://archive.hxmt.cn/proposal}{\textit{Insight}-HXMT} for this work.


\begin{table}
\centering
\caption{Evolution of QPO parameters, Time-lags in the $4-10$, $10-30$, and $30-150$\,keV bands relative to the $2-4$\,keV band, and best fitted POS model parameters.}
\label{tab:table_qpo}
\small
\begin{tabular}{cccccccccc} 
 \hline\hline
\textbf{ObsID}$^a$ & \textbf{Day} & {\boldmath$\nu_{QPO}$}$^b$ & \textbf{RMS}$^b$ & {\boldmath$\Gamma$}$^c$ & \textbf{TLag}$_{(4-10\,keV)}$ & \textbf{TLag}$_{(10-30\,keV)}$ & \textbf{TLag}$_{(30-150\,keV)}$ & {\boldmath$X_s$}$^d$ & {\boldmath$R$}$^d$ \\
   & (MJD) &  (Hz)  &  (\%)   &     &  (s)  &  (s)  & (s)  &  ($r_s$)  & \\  
   (1)  &  (2)  &  (3) & (4) & (5) & (6)  & (7)  &  (8) &  (9)   &   (10)   \\
\hline
X0101 & 60181.4195 & $0.088^{\pm0.003}$ &  $15.2^{\pm2.0}$  &   $1.744^{\pm0.023}$  &  $ 0.096^{\pm0.034}$ & $ 0.056^{\pm0.090}$  &         ***          &  773  &  4.00  \\ 
X0106 & 60182.0925 & $0.177^{\pm0.006}$ &  $15.9^{\pm4.3}$  &   $1.816^{\pm0.122}$  &  $ 0.022^{\pm0.017}$ & $ 0.033^{\pm0.033}$  & $ 0.126^{\pm0.070}$ &  585  &  3.96  \\ 
X0201 & 60183.1313 & $0.433^{\pm0.005}$ &  $18.8^{\pm2.1}$  &   $2.082^{\pm0.116}$  &  $ 0.035^{\pm0.007}$ & $ 0.055^{\pm0.012}$  &         ***          &  368  &  3.74  \\ 
X0208 & 60184.0746 & $0.627^{\pm0.006}$ &  $17.6^{\pm3.9}$  &   $2.124^{\pm0.065}$  &  $ 0.020^{\pm0.005}$ & $ 0.031^{\pm0.008}$  & $ 0.056^{\pm0.025}$ &  250  &  3.42  \\ 
X0301 & 60185.3848 & $0.890^{\pm0.004}$ &  $19.7^{\pm1.2}$  &   $2.192^{\pm0.039}$  &  $ 0.014^{\pm0.001}$ & $ 0.022^{\pm0.002}$  & $ 0.023^{\pm0.005}$ &  209  &  2.90  \\ 
X0307 & 60186.1854 & $1.066^{\pm0.004}$ &  $19.5^{\pm1.2}$  &   $2.263^{\pm0.038}$  &  $ 0.010^{\pm0.001}$ & $ 0.012^{\pm0.001}$  &         ***          &  181  &  2.90  \\ 
X0314 & 60187.1033 & $1.087^{\pm0.006}$ &  $18.6^{\pm1.9}$  &   $2.155^{\pm0.048}$  &  $ 0.010^{\pm0.001}$ & $ 0.010^{\pm0.002}$  &         ***          &  179  &  2.90  \\ 
X0408 & 60188.1673 & $1.331^{\pm0.004}$ &  $18.8^{\pm1.0}$  &   $2.176^{\pm0.041}$  &  $ 0.007^{\pm0.001}$ & $ 0.006^{\pm0.001}$  &         ***          &  177  &  2.89  \\ 
X0501 & 60189.1485 & $1.308^{\pm0.005}$ &  $19.1^{\pm1.1}$  &   $2.193^{\pm0.029}$  &  $ 0.007^{\pm0.001}$ & $ 0.007^{\pm0.001}$  & $ 0.009^{\pm0.072}$ &  175  &  2.88  \\ 
X0508 & 60190.0870 & $1.446^{\pm0.007}$ &  $18.9^{\pm1.2}$  &   $2.270^{\pm0.019}$  &  $ 0.005^{\pm0.001}$ & $ 0.006^{\pm0.001}$  & $ 0.004^{\pm0.003}$ &  173  &  2.87  \\ 
X0601 & 60191.1311 & $1.254^{\pm0.006}$ &  $18.6^{\pm1.0}$  &   $2.196^{\pm0.016}$  &  $ 0.008^{\pm0.001}$ & $ 0.011^{\pm0.002}$  & $ 0.016^{\pm0.003}$ &  171  &  2.85  \\ 
X0608 & 60192.0692 & $1.121^{\pm0.005}$ &  $17.6^{\pm1.0}$  &   $2.132^{\pm0.032}$  &  $ 0.009^{\pm0.001}$ & $ 0.012^{\pm0.002}$  & $ -0.01^{\pm 0.08}$ &  169  &  2.83  \\ 
X0616 & 60193.1265 & $1.478^{\pm0.006}$ &  $17.6^{\pm0.9}$  &   $2.236^{\pm0.016}$  &  $ 0.006^{\pm0.001}$ & $ 0.005^{\pm0.001}$  & $ 0.007^{\pm0.002}$ &  167  &  2.80  \\ 
X0801 & 60194.1048 & $1.295^{\pm0.007}$ &  $17.9^{\pm0.9}$  &   $2.179^{\pm0.131}$  &  $ 0.008^{\pm0.001}$ & $ 0.007^{\pm0.002}$  & $ 0.023^{\pm0.003}$ &  165  &  2.78  \\ 
X0901 & 60195.1635 & $1.317^{\pm0.006}$ &  $17.3^{\pm1.0}$  &   $2.180^{\pm0.026}$  &  $ 0.007^{\pm0.001}$ & $ 0.010^{\pm0.002}$  & $ 0.017^{\pm0.004}$ &  163  &  2.75  \\ 
X1001 & 60196.1546 & $1.384^{\pm0.005}$ &  $16.5^{\pm1.0}$  &   $2.176^{\pm0.021}$  &  $ 0.007^{\pm0.001}$ & $ 0.006^{\pm0.002}$  & $ 0.008^{\pm0.004}$ &  161  &  2.72  \\ 
X1101 & 60197.1456 & $1.457^{\pm0.005}$ &  $15.9^{\pm1.1}$  &   $2.135^{\pm0.031}$  &  $ 0.006^{\pm0.001}$ & $ 0.007^{\pm0.002}$  & $ 0.009^{\pm0.003}$ &  152  &  2.65  \\ 
X1201 & 60198.1366 & $1.795^{\pm0.008}$ &  $16.3^{\pm0.9}$  &   $2.213^{\pm0.026}$  &  $ 0.004^{\pm0.002}$ & $ 0.002^{\pm0.002}$  & $ 0.008^{\pm0.004}$ &  141  &  2.49  \\ 
X1301 & 60199.1275 & $2.349^{\pm0.008}$ &  $16.2^{\pm0.9}$  &   $2.335^{\pm0.084}$  &  $ 0.001^{\pm0.002}$ & $-0.003^{\pm0.002}$  & $-0.001^{\pm0.004}$ &  130  &  2.26  \\ 
X1401 & 60200.1185 & $2.944^{\pm0.027}$ &  $17.8^{\pm0.9}$  &   $2.353^{\pm0.096}$  &  $-0.002^{\pm0.001}$ & $-0.008^{\pm0.002}$  & $-0.008^{\pm0.003}$ &  119  &  1.99  \\ 
X1402 & 60200.1880 & $2.835^{\pm0.012}$ &  $16.2^{\pm1.0}$  &   $2.363^{\pm0.073}$  &  $ 0.000^{\pm0.001}$ & $-0.008^{\pm0.001}$  & $-0.014^{\pm0.002}$ &  122  &  2.00  \\ 
X1403 & 60200.3298 & $2.619^{\pm0.011}$ &  $16.5^{\pm0.8}$  &   $2.367^{\pm0.076}$  &  $ 0.000^{\pm0.001}$ & $-0.006^{\pm0.001}$  & $-0.008^{\pm0.002}$ &  127  &  2.00  \\ 
X1404 & 60200.4620 & $2.642^{\pm0.010}$ &  $16.4^{\pm0.8}$  &   $2.432^{\pm0.063}$  &  $-0.000^{\pm0.001}$ & $-0.006^{\pm0.001}$  & $-0.010^{\pm0.002}$ &  131  &  2.00  \\ 
X1405 & 60200.5941 & $2.547^{\pm0.017}$ &  $16.1^{\pm1.1}$  &   $2.432^{\pm0.100}$  &  $ 0.000^{\pm0.002}$ & $-0.004^{\pm0.002}$  & $-0.006^{\pm0.003}$ &  136  &  2.00  \\ 
X1406 & 60200.7263 & $2.157^{\pm0.012}$ &  $16.3^{\pm1.2}$  &   $2.344^{\pm0.049}$  &  $ 0.001^{\pm0.002}$ & $-0.002^{\pm0.002}$  &         ***          &  140  &  2.00  \\ 
X1407 & 60200.8585 & $2.069^{\pm0.008}$ &  $16.1^{\pm0.9}$  &   $2.341^{\pm0.035}$  &  $ 0.002^{\pm0.002}$ & $-0.000^{\pm0.002}$  & $ 0.000^{\pm0.003}$ &  144  &  2.00  \\ 
X1408 & 60200.9907 & $2.052^{\pm0.008}$ &  $16.3^{\pm0.8}$  &   $2.287^{\pm0.032}$  &  $ 0.002^{\pm0.001}$ & $ 0.001^{\pm0.002}$  & $-0.000^{\pm0.003}$ &  149  &  2.00  \\ 
X1501 & 60201.2468 & $1.924^{\pm0.008}$ &  $16.1^{\pm0.9}$  &   $2.307^{\pm0.034}$  &  $ 0.002^{\pm0.004}$ & $ 0.001^{\pm0.004}$  & $ 0.001^{\pm0.006}$ &  157  &  2.00  \\ 
X1701 & 60203.2293 & $2.647^{\pm0.015}$ &  $16.2^{\pm1.0}$  &   $2.321^{\pm0.101}$  &  $-0.001^{\pm0.002}$ & $-0.005^{\pm0.002}$  & $-0.006^{\pm0.003}$ &  145  &  1.74  \\ 
X1801 & 60204.2205 & $2.598^{\pm0.008}$ &  $15.4^{\pm0.8}$  &   $2.278^{\pm0.059}$  &  $-0.000^{\pm0.001}$ & $-0.006^{\pm0.001}$  & $-0.010^{\pm0.002}$ &  133  &  1.50  \\ 
X1901 & 60205.1457 & $4.353^{\pm0.013}$ &  $15.6^{\pm0.8}$  &   $2.475^{\pm0.034}$  &  $-0.002^{\pm0.001}$ & $-0.008^{\pm0.001}$  & $-0.014^{\pm0.002}$ &  118  &  1.27  \\ 
X2002 & 60206.2050 & $7.496^{\pm0.051}$ &  $16.8^{\pm0.5}$  &   $2.606^{\pm0.029}$  &  $-0.003^{\pm0.001}$ & $-0.002^{\pm0.001}$  & $ 0.001^{\pm0.002}$ &   96  &  1.04  \\ 
X2003 & 60206.2778 & $8.782^{\pm0.026}$ &  $15.0^{\pm0.6}$  &   $2.622^{\pm0.031}$  &  $-0.005^{\pm0.001}$ & $-0.004^{\pm0.001}$  & $ 0.003^{\pm0.001}$ &   94  &  1.03  \\ 
X2004 & 60206.4099 & $5.592^{\pm0.040}$ &  $10.3^{\pm1.3}$  &   $2.644^{\pm0.040}$  &  $-0.004^{\pm0.001}$ & $-0.008^{\pm0.001}$  & $-0.011^{\pm0.002}$ &   91  &  0.99  \\
X2101 & 60207.1282 & $5.321^{\pm0.018}$ &  $15.3^{\pm0.7}$  &   $2.662^{\pm0.034}$  &  $-0.003^{\pm0.001}$ & $-0.009^{\pm0.001}$  & $-0.014^{\pm0.002}$ &  ***  &   ***   \\
X2201 & 60208.1194 & $5.246^{\pm0.018}$ &  $15.7^{\pm0.8}$  &   $2.654^{\pm0.032}$  &  $-0.003^{\pm0.001}$ & $-0.008^{\pm0.001}$  & $-0.015^{\pm0.002}$ &  ***  &   ***   \\
X2301 & 60209.1767 & $3.593^{\pm0.016}$ &  $16.0^{\pm0.7}$  &   $2.457^{\pm0.028}$  &  $-0.001^{\pm0.001}$ & $-0.006^{\pm0.001}$  & $-0.012^{\pm0.001}$ &  ***  &   ***   \\
X2401 & 60210.1679 & $3.599^{\pm0.014}$ &  $15.6^{\pm0.7}$  &   $2.472^{\pm0.025}$  &  $-0.001^{\pm0.001}$ & $-0.007^{\pm0.001}$  & $-0.011^{\pm0.001}$ &  ***  &   ***   \\
X2501 & 60211.1591 & $5.836^{\pm0.024}$ &  $16.3^{\pm0.7}$  &   $2.672^{\pm0.032}$  &  $-0.002^{\pm0.001}$ & $-0.004^{\pm0.001}$  & $-0.007^{\pm0.001}$ &  ***  &   ***   \\
X2601 & 60212.1504 & $6.629^{\pm0.032}$ &  $16.1^{\pm0.7}$  &   $2.733^{\pm0.033}$  &  $-0.002^{\pm0.001}$ & $-0.001^{\pm0.001}$  & $ 0.005^{\pm0.006}$ &  ***  &   ***   \\
X2701 & 60213.1417 & $7.236^{\pm0.071}$ &  $17.9^{\pm0.7}$  &   $2.718^{\pm0.034}$  &  $-0.002^{\pm0.001}$ & $-0.001^{\pm0.002}$  & $ 0.005^{\pm0.013}$ &  ***  &   ***   \\
X2802 & 60214.2061 & $4.985^{\pm0.030}$ &  $16.0^{\pm0.8}$  &   $2.671^{\pm0.036}$  &  $-0.002^{\pm0.001}$ & $-0.004^{\pm0.001}$  & $-0.009^{\pm0.002}$ &  ***  &   ***   \\
X2901 & 60215.1242 & $3.941^{\pm0.016}$ &  $15.1^{\pm0.8}$  &   $2.499^{\pm0.059}$  &  $-0.001^{\pm0.001}$ & $-0.003^{\pm0.002}$  & $-0.004^{\pm0.002}$ &  ***  &   ***   \\
X3001 & 60216.1155 & $6.192^{\pm0.044}$ &  $16.9^{\pm0.8}$  &   $2.679^{\pm0.042}$  &  $-0.002^{\pm0.001}$ & $-0.001^{\pm0.002}$  & $ 0.009^{\pm0.008}$ &  ***  &   ***   \\
X3101 & 60217.1729 & $5.403^{\pm0.028}$ &  $15.8^{\pm0.8}$  &   $2.645^{\pm0.038}$  &  $-0.001^{\pm0.001}$ & $-0.001^{\pm0.001}$  & $ 0.010^{\pm0.014}$ &  ***  &   ***   \\
X3201 & 60218.1641 & $6.521^{\pm0.040}$ &  $16.4^{\pm0.8}$  &   $2.679^{\pm0.043}$  &  $-0.001^{\pm0.001}$ & $ 0.001^{\pm0.002}$  & $ 0.005^{\pm0.002}$ &  ***  &   ***   \\
X3301 & 60219.1552 & $5.755^{\pm0.034}$ &  $16.7^{\pm0.7}$  &   $2.659^{\pm0.041}$  &  $-0.001^{\pm0.001}$ & $-0.000^{\pm0.002}$  & $-0.004^{\pm0.002}$ &  ***  &   ***   \\
X3308 & 60220.0850 & $7.077^{\pm0.060}$ &  $16.9^{\pm0.9}$  &   $2.704^{\pm0.047}$  &  $-0.000^{\pm0.002}$ & $ 0.001^{\pm0.002}$  & $ 0.003^{\pm0.003}$ &  ***  &   ***   \\
X3401 & 60221.1373 & $8.988^{\pm0.057}$ &  $15.9^{\pm1.1}$  &   $2.703^{\pm0.081}$  &  $-0.001^{\pm0.003}$ & $-0.004^{\pm0.003}$  & $ 0.001^{\pm0.003}$ &  ***  &   ***   \\
X3502 & 60222.1988 & $8.414^{\pm0.062}$ &  $16.4^{\pm0.9}$  &   $2.717^{\pm0.069}$  &  $-0.002^{\pm0.002}$ & $ 0.001^{\pm0.002}$  & $ 0.004^{\pm0.003}$ &  ***  &   ***   \\
\hline
\end{tabular}
\leftline{\textbf{Notes.} Superscripts on the parameter values represent average errors obtained using the \texttt{err} task in \texttt{XSPEC} with 90\% conﬁdence.}
\leftline{$^a$ X(=P061433803) denotes the initial part of \textit{Insight}-HXMT Observation ID. }
\leftline{$^b$ Lorentzian model fitted LFQPO frequency ($\nu_{QPO}$) and fractional root mean square amplitude (RMS) are shown in columns (3)-(4).}
\leftline{$^c$ $\Gamma$ denotes the photon index.}
\leftline{$^d$ $X_s$ and $R$ denotes the POS model fitted shock location and shock compression ratio. $X_s$ is measured in units of the Schwarzschild radius, $r_s = 2GM/c^2$.}
\leftline{($R = \rho_+/\rho_-$, ratio of the post-shock to the pre-shock matter density).}
\leftline{Entries marked `***' indicate cases where (i) time lags could not be measured for these observations (Column 8), or }
\leftline{(ii) POS model fitting parameters ($X_s$, $R$) were not available for these observations in Paper-II (Columns 9,10).}
\end{table}


\bsp	
\label{lastpage}
\end{document}